\documentclass[useAMS,usenatbib]{mn2e}
\usepackage{graphicx}    
\DeclareGraphicsExtensions{.ps,.pdf,.png}
\makeatletter
\def\fps@figure{htbp}
\makeatother
\def \aap  {A\&A}
\def \aaps  {A\&AS}

\def \apj  {ApJ}
\def \apjs  {ApJS}
\def \araa  {ARA\&A}

\def \mnras {MNRAS}

\def\apjl {ApJL}
\newcommand*{\Msun}{\ensuremath{\mathrm{M_\odot}}}%
\begin{document}
\title[Galaxy Stellar Mass Functions in the GNS]{A Deep Probe of the Galaxy Stellar Mass Functions at $z\sim 1-3$ with the GOODS NICMOS Survey}
\author[Mortlock et al.]{Alice~Mortlock$^{1}$, Christopher~J.~Conselice$^1$, Asa~F.~L.~Bluck$^{1,2}$, Amanda~E.~Bauer$^{1,3}$, 
\newauthor Ruth~Gr\"utzbauch$^{1}$, Fernando~Buitrago$^{1}$, Jamie~Ownsworth$^{1}$  \footnotemark[0]\\
$^{1}$University of Nottingham, School of Physics and Astronomy, Nottingham, NG7 2RD UK \\ $^{2}$ Gemini Observatory, Hilo, Hawaii 96720, USA \\ $^{3}$Anglo-Australian Observatory, PO Box 296, Epping, NSW 2111, Australia}

\date{Accepted ??. Received ??; in original form ??}
\pagerange{\pageref{firstpage}--\pageref{lastpage}} \pubyear{2002}
\maketitle

\label{firstpage}

\begin{abstract}
We use a sample of 8298 galaxies observed as part of the HST $H_{160}$-band GOODS NICMOS Survey (GNS) to construct the galaxy stellar mass function both as a function of redshift and stellar mass up to $z = 3.5$. Our mass functions are constructed within the redshift range $z=1-3.5$
and consist of galaxies with stellar masses of $M_{*}=10^{12} M_{\odot}$ down to nearly dwarf galaxy masses of $M_{*}=10^{8.5} M_{\odot}$ in the lowest redshift bin.
We discover that a significant fraction of all massive $M_{*}>10^{11} M_{\odot}$ galaxies are in place up to the highest redshifts we probe, with a decreasing fraction of lower mass galaxies present at all redshifts.  This is an example of `galaxy mass downsizing', and is the result of massive galaxies forming before lower mass ones, and not just simply ending their star formation earlier as in traditional downsizing
scenarios, whose effect is seen at $z<1.5$. By fitting Schechter functions to our mass functions we find that the faint end slope ranges from $\alpha=-1.36$ to $-1.73$, which is significantly steeper than what is found in previous investigations of the mass function at high redshift. We demonstrate that this steeper mass function better matches the stellar mass added due to star formation, thereby alleviating some of the mismatch between these two measures of the evolution of galaxy mass. We furthermore examine the stellar mass function divided into blue/red systems, as well as for star forming and non-star forming galaxies.  We find a similar mass downsizing present for both blue/red and star-forming/non-star forming galaxies, and further find that red galaxies dominate at the high mass end of the mass function, but that the low mass galaxies are mostly all blue, and therefore blue galaxies are creating the steep mass functions observed at $z > 2$.  We furthermore show that, although there is a downsizing such that high mass galaxies are nearer their $z=0$ values at high redshift, this turns over at masses $M_{*}\sim10^{10} M_{\odot}$, such that the lowest mass galaxies are more common than galaxies at slight higher masses, creating a `dip' in the observed galaxy mass function.  We argue that the galaxy assembly process may be driven by different mechanisms at low and high masses, and that the efficiency of the galaxy formation process is lowest at masses $M_{*}\sim10^{10} M_{\odot}$ at $1 < z < 3$. Finally, we calculate the integrated stellar mass density for the total, blue and red populations. We find the integrated stellar mass density of the total and blue galaxy population is consistent with being constant over $z=1-2$, while the red population shows an increase in integrated stellar mass density over the same redshift range.
\end{abstract}


\section{Introduction}
A deep understanding of the high redshift universe is vital in order to
complete our knowledge of galaxy formation, and hence uncover the history
of the universe as a whole. With the continued development of
instrumentation and of new analysis techniques our knowledge of the high
redshift universe is increasing rapidly. Thanks to new deep imaging and
multiobject spectroscopy we now have the power to routinely look back at
the universe over cosmic time to witness step by step evolution. The
result of this has been a wealth of observations of large samples of
galaxies over a large redshift range in various surveys, such as within GOODS
(\citealt{Giav04}), COSMOS and z-COSMOS (\citealt{Scov07} and
\citealt{Lill07}) and AEGIS (\citealt{Davi07}). These large samples give
us the power to achieve statistically meaningful results concerning the
evolution of galaxy properties.

As a result a detailed picture is beginning to form regarding when
galaxy stellar mass is built up over cosmic time. Various studies
(\citealt{Dick03}, \citealt{Dror05}, \citealt{Cons07}, \citealt{Elsn08},
\citealt{Pere08}) have focused on the change of stellar mass density
with time, and have seen generally consistent results. These studies find that
the integrated stellar mass density decreases at higher redshift, as
expected since the
ongoing process of star formation increases the amount of stellar mass
in the universe over time. These studies also show that roughly 50\%
of the mass density of the universe is in place by $z \sim 1$. This
implies prior to this redshift the comoving stellar mass density has a
more rapid evolution. This ties in with studies that show that the star
formation rate peak is at $z\ge1-2$ (\citealt{Mada96}, \citealt{Hopk04},
\citealt{Hopk06}).

While studies of the star formation history has been the traditional way to
examine and probe galaxy evolution, a great deal of research has been
carried out investigating the evolution of
galaxies using the galaxy stellar mass function.   Early examples of the
measurement of the local galaxy mass function at $z \sim 0$
have been carried out by e.g. \citet{Cole01} and \citet{Bell03}, using surveys
such as 2MASS, 2dF, and the Sloan Digital Sky Survey.
data. These and other investigations construct the stellar mass function,
and integrated stellar mass density within the local universe, and
hence provide a vital benchmark for comparison with higher redshifts.

It is important to understand the stellar mass function of the
local universe in itself as this traces the integrated star formation and mass assembly
history over
the entire universe.  However, by extending similar studies to higher
redshift we can
investigate not only galaxy growth as a function of stellar mass, but
also the growth of integrated galaxy stellar mass with time. That is,
we can trace the evolution of stellar mass for galaxies of different
stellar masses, i.e., very massive galaxies vs. lower mass galaxies,
over time.   This gives us insights into the growth of galaxies of different
stellar masses due to star formation, mergers and other assembly processes.
Many studies have investigated the stellar mass function up to
$z=2$ using relatively large area surveys (\citealt{Font04}, \citealt{Bund06},
\citealt{Borc06}, \citealt{Fran06} \citealt{Bell07}, \citealt{Bolz09}).
This has been further extended to even higher redshifts using deeper surveys,
generally within a much smaller area (\citealt{Cons05}, \citealt{Font06},
\citealt{Pere08}, \citealt{Kaji09}).

These studies of the stellar mass function at high redshifts allows us to form
a picture of the high mass end of the stellar mass function as a
function of redshift.
It is widely agreed that the dominance of star formation within massive galaxies
ends much earlier than within low stellar mass galaxies by $z \sim 1$ (\citealt{Bund06}). This is one
form of galaxy downsizing, whereby the higher stellar mass galaxies have
their star formation truncated, or gas depleted, earlier than lower mass
galaxies. This
formation scenario was first observed by \citet{Cow96} and subsequently
observed in various studies including e.g. \citet{Baue05}, \citet{Feul05},
\citet{Bund06}, and \citet{Verg08}. Downsizing is now accepted as part
of the formation scenario of galaxies, but is not yet fully understood.  What is
also not fully understood is whether the related, but different process of mass
downsizing, is occurring in tandem, such that the high mass galaxies form their
stellar mass earlier than lower mass galaxies, and when this stellar mass differentiated
galaxy formation process first reveals itself.

Despite the considerable work done investigating the high mass end of
the stellar mass function, there are still many issues that are not yet
fully understood. Firstly, the generally shape of the high redshift stellar mass
function is not well described. Nearby stellar mass functions are well
fit by a form of the Schechter function, and there are various
investigations regarding how the parameters of this fit change with
redshift. Secondly, difficulties with obtaining deep data at high
redshifts mean that the low mass end of the stellar mass function has
not been explored as fully, or to as high a redshift. More recent work
has started to uncover a possible steepening with redshift of the low
stellar mass end, and a ``dip" in the intermediate stellar mass range (e.g. \citealt{Kaji09}). It
has been suggested that this is a result of evolution of different
galaxy populations driven by their mass (\citealt{Bolz09}, \citealt{Dror09},
\citealt{Ilbe09}, \citealt{Pozz09}). The exact nature and reasons behind
such features of the stellar mass function are not well understood, and
this can only be improved upon with deeper, more robust data. This will
then lead to a better understanding of how the populations of galaxies,
as defined by stellar mass, change and evolve over time.

In this paper we use data from the GOODS NICMOS survey (GNS) to
investigate how the stellar mass function evolves from $z= 1$ to $z \sim
3.5$. By examining the stellar mass functions of galaxies ranging in
stellar mass from $M_{*}=10^{12} M_{\odot}$ to as low as $M_{*}=10^{8.5}
M_{\odot}$ we investigate when, and which galaxies are forming at
various epochs in the universe.  The depth of the GNS data is such that
we are able to probe over a factor of 10$^{3}$ the stellar mass evolution
up to $z \sim 3.5$ and importantly trace how galaxies of different masses
are evolving with time.   We find throughout this paper a differential in the
stellar mass function and how it evolves, revealing a strong stellar mass
dependence in the galaxy formation process.  We describe this and give
some general explanations for how this differential evolution can occur due
to different physical processes.

The paper is set out as follows: Section \ref{sec:GNS} discusses the
GOODS-NICMOS Survey, the galaxy sample and how the data used in this
paper was obtained. Section \ref{sec:totalMF} examines the galaxy
stellar mass functions of all the galaxies in various redshift bins. In Sections \ref{sec:MFforBR} and \ref{sec:MFforSF} we split the galaxies into
blue and red and star forming and non-star forming respectively. Section
\ref{sec:massden} describes the calculation of the stellar mass
densities for the total sample and for the red and blue populations.
Sections \ref{sec:diss} and \ref{sec:summ} contain the discussion and
summary of our findings respectively. Throughout this paper we assume
$\Omega_{M}=0.3$, $\Omega_{\Lambda}=0.7$ and $H_{0}=70$ km s$^{-1}$
Mpc$^{-1}$. AB magnitudes and a Salpeter IMF are
used throughout.

\section{Data reduction and the GNS}
\label{sec:GNS}
\subsection{The GOODS-NICMOS Survey}
The galaxy sample used in this work is taken from and imaged as part of the GOODS-NICMOS Survey (GNS) (\citealt{Cons10}). The GNS consists of 60 pointings of the HST NICMOS-3 camera utilising a total of 180 orbits. The field of view of NICMOS-3 is 51.2 arcsec x 51.2 arcsec with a pixel scale of 0.203 arcsec/pixel. Each NIC3 tile was observed in six exposures, which combined give a pixel scale of 0.1 arcsec/pixel with a point spread function (PSF) of $\sim$ 0.3 arcsec full width half maximum (FWHM). Within the GNS we find 8298 galaxies in the $H_{160}$ band (F160W), and the 60 pointings are designed to contain as many massive galaxies ($M_{*}>10^{11}M_{\sun}$) as possible. These massive galaxies are in the redshift range $z=1.7-2.9$ and at a depth of three orbits. The galaxies are redshift selected based on their optical to infrared colour described in \citet{Cons10}. Details on the data reduction pipeline are described in \citet{Mage07}. The detections and photometry were done using SExtractor (\citealt{Bert96}). At 5$\sigma$ the limiting magnitude is $H_{160}$=26.8, a marked improvement on the GOODS ground based data at $K=$24.5 (\citealt{Retz09}). Further description of the GNS, the pointings and the target selection is given in \citet{Cons10}. Other analysis of the GNS data set can be found in \citet{Buit08}, \citet{Bluc09}, Bauer et al. (2010, submitted), \citet{Bluc10}, \citet{Grut10}.

\subsection{Photometric Redshifts}
Thanks to the large amount of optical data covering the GOODS fields, the $H_{160}$ band sources are matched to a catalogue of $B, V, i$ and $z$ band data. This photometry is avaliable down to a limiting magnitude of B$\sim$28.2, and the matching is done within a radius of 2$''$. However the mean separation between the optical and $H_{160}$ band coordinates is markedly better than this with $\langle r\rangle$ $\sim$0.28$\pm$0.4", which is roughly the resolution of NICMOS. With this multiband data template spectra we fit to the $BVizH$ photometric data points. This was handled in two different ways to overcome the degeneracy in colour-redshift space.

The first uses HYPERZ (\citealt{Bolz09}) which is the standard $\chi^{2}$ minimisation technique. HYPERZ uses model spectra which were constructed using the evolutionary codes of \citet{Bruz93}. Here we use five evolutionary types, E, Sa, Sc, Im and a single burst scenario. The reddening law is that of \citet{Calz00}. The most likely redshift is then computed in the age, metalicity, reddening parameter space, giving the best fit redshift, corresponding probability and several other best fit parameteres.

The second method used to obtain our photometric redshifts is the Bayesian approach using BPZ. This is a similar fitting method to HYPERZ but employs empirical SEDs rather than model ones. As well as this, BPZ uses the maximum likelihood in the same parameter space as HYPERZ, but with the addition of empirical information regarding the likelihood of a certain combination of parameters. This is known as prior information, or priors. Here, we used the distribution of magnitudes of different galaxy types as a function of redshift as the priors (from HDF-N, \citealt{Beni00}). Therefore in this case, not only does the code find the best fit redshift solution and spectral type, it also consideres how likely it is to find a galaxy of that type and magnitude at that redshift.

A total of 906 spectroscopic redshifts are avaliable for comparison with our photometric redshifts from the GOODS-N (\citealt{Barg08}) and the GOODS-S (\citealt{Wuyt08}) fields. The reliability of photometric redshifts is defined as $\Delta z/(1+z)=(z_{spec}-z_{phot})/(1+z_{spec})$, the median error ($\langle \Delta z/(1+z) \rangle$) and the r.m.s ($\sigma$) are as follows. For HYPERZ, the GOODS-N gives $\langle \Delta z/(1+z) \rangle$ = 0.027 with $\sigma$=0.04, and for the GOODS-S $\langle \Delta z/(1+z) \rangle$ = 0.043 with $\sigma$=0.04. For BPZ, the GOODS-N $\langle \Delta z/(1+z) \rangle$ = 0.07 with $\sigma$=0.05 and for the GOODS-S $\langle \Delta z/(1+z) \rangle$ = 0.07 with $\sigma$=0.06. In both codes the extreme catastrophic outliers, as defined by $|\Delta z/(1+z)|>0.5$, are around $\sim$6\%. The dependence of the redshift on the $H_{160}$ magnitude shows the reliability of redshift within our selection methods and in this case HYPERZ shows a slightly better performance (\citealt{Grut10}). Our work shows good agreement with photometric redshifts from past surveys (e.g. FIREWORKS), although our sample is bright, and it is unclear if this would be the case for fainter galaxies.

We further investigate the performance of HYPERZ at different redshifts,
at low redshift ($z<1.5$) and in the redshift range of $1.5 \leq z
\leq 3$, which is the redshift range of the galaxy sample we use in
this study.  For the high redshift sample we obtain an average offset
$\langle \Delta z/(1+z) \rangle = 0.06$ and a RMS of $\sigma_{\Delta z
/(1+z)} = 0.10$, with a fraction of catastrophic outliers of $20\%$.
Here catastrophic outliers are defined as galaxies with $|\Delta
z/(1+z)| > 0.3$, which corresponds to $\sim$ 3 times the RMS scatter. Galaxies below $z=1.5$ show a slightly lower, but still comparable scatter of $\sigma_{\Delta z /(1+z)} = 0.08$, however the outlier fraction decreases dramatically to only $\sim 2\%$. Furthermore, we simulate the effects of the photometric redshift errors on our results throughout this paper.

\subsection{Stellar Masses and Colours}
The stellar masses we use are obtained from the same $BVizH$ catalogue as used to measure the photometric redshifts described previously, using a standard multicolour stellar population fitting technique (e.g., \citealt{Cons08} and \citealt{Cons10}). The photometry is fit to different star formation histories, based on the redshift of the galaxy, with spectroscopic redshifts used when avaliable. This produces a distribution of likely stellar masses, rest frame optical colours and various other parameters based on a Baysian approach. The model SEDs are constructed from \citet{Bruz03} models, and the star formation history is characterised by an exponentially declining model with various ages, metalicities and dust extinctions. The star formation rate is parameterised by an e-folding time and an age such that,
\begin{equation}
SFR(t) \sim SFR_{0}\times e ^{-\frac{t}{\tau}}.
\label{eq:SFRexp}
\end{equation}
The parameters in Equation \ref{eq:SFRexp} are varied randomly within the ranges; $\tau$=0.01 to 10 Gyrs, t=0 to 10 Gyrs, metallicity= 0.0001 to 0.05 and the dust content is parameterised by $\tau_{v}$=0.0, 0.5, 1, 2. These model SEDs are then fit to the observed photometric data points using a Bayesian approach resulting in a likelihood distribution of stellar mass, age and absolute magnitude for each possible star formation history. The stellar mass is determined based on this distribution, where the most likely stellar mass produces a peak in the distribution, and the uncertainty is the width. The final errors produced are a result of the models used and are found to be in the range 0.2 to 0.3 dex. While several of the other parameters produced by this method are not reliable, the stellar masses and colours are robust (see \citealt{Bund06}  and \citealt{Cons07} for further explanation.). It is also possible that the stellar masses are an over estimate due to the poor treatment of the TP-AGB phase in a star's life. The effects of this are much less important at the rest frame wavelengths used in this study, especially in the infrared H-band. By using the newer models of Bruzual and
Charlot (2010) which include a more proper treatment of TP-AGB stars we find it only lowers the masses in our massive galaxy sample by $<$0.07 dex. This effect is much smaller than the stellar mass error and the effects of cosmic variance and it is therefore considered negligible in this work.

\subsection{Star Formation Rates}
\label{sec:SFRexplan}
To obtain star formation rates (SFRs) for our GNS galaxies we use the rest-frame ultraviolet (UV) luminosity, which we derive from observed rest frame optical light following the procedure described in Bauer et al. (2010, submitted).  The UV luminosity is closely related to the level of ongoing star formation because it is mainly produced by short-lived O and B stars.  An advantage of using UV light to estimate star formation is that it remains largely unaffected by the age-metallicity degeneracy (\citealt{Wort94}), but as is well known, star formation rate estimates derived from UV light are strongly affected by dust extinction.

The 2800$\mathrm{\AA}$ rest-frame luminosity is calculated from the observed optical Hubble/ACS $z$-band flux density, which corresponds to rest-frame wavelengths of 3400 - 2125$\mathrm{\AA}$ for $z = 1.5 - 3$ galaxies. To derive SFR$_{UV}$ we apply a simple K-correction derived from the redshift of each object as log($1+z$) (\citealt{Kim96}, \citealt{Dadd04}) and use the \citet{Kenn98} conversion from 2800$\mathrm{\AA}$ luminosity to star formation rate assuming a Salpeter IMF:

\begin{equation} \label{eq:sfr}
\textrm{SFR}_{\mathrm{UV}}\,(\Msun \,\textrm{yr}^{-1}) = 1.4 \times10^{-28}\, L_\nu\,(\textrm{ergs }\,\textrm{s}^{-1}\,\textrm{Hz}^{-1}).
\end{equation}
Before dust extinction is taken into account (Section \ref{dust}), we find at $z=1.5$ a limiting SFR$_{UV}= 0.28\pm0.1~\Msun$yr$^{-1}$ and at $z=3.0$, we find a limit of SFR$_{UV}= 0.98\pm0.3~ \Msun$yr$^{-1}$.  The errors take into account photometric errors and the error in the conversion from a luminosity to a star formation rate (see also Bauer et al. (2011, in prep.).

\subsection{Dust Correction}
\label{dust}
To obtain reliable star formation rates from the rest-frame ultraviolet, we need to account for the amount of light obscured by dust, which is a non-trivial problem. \citet{Meur99} demonstrated a correlation between dust attenuation and rest-frame UV slope, $\beta$, for a sample of nearby starburst galaxies (where $F_\lambda \sim \lambda^{\beta}$). Updated studies of local galaxies using the {\it Galaxy Evolution Explorer} (GALEX) near-ultraviolet band and $z\sim2$ galaxies show that the UV slope from the local starburst relation can be used to recover the dust attenuation of a vast majority of moderately luminous galaxies at $z\sim2$ (\citealt{Buat05}, \citealt{Seib05}, \citealt{Redd10}).

To determine the UV slope, we use the SED-fitting procedure.  We  fit a spectral energy distribution (SED) to multi-wavelength observations from optical to mid-infrared, following the procedure described in \citet{Pere08} and Bauer et al. (2011).   Briefly, the UV through MIR SEDs obtained for all sources in the GOODS fields were fitted with stellar population synthesis models. Then, the best-fitting templates were used to get synthetic estimations of the UV emission at 1600$\mathrm{\AA}$ and 2800$\mathrm{\AA}$.  From the synthetic, model-derived UV luminosities at 1600$\mathrm{\AA}$ and 2800$\mathrm{\AA}$, we calculate the spectral slope, $\beta$.  We use the \citet{Calz00} law to derive A$_{2800}$ from the UV spectral slope, which we then apply to the UV-derived star formation rates. Using this method we find an average extinction value of A$_{2800}=1.6~\pm$~1.2 mag.  We find that all galaxies in the GNS exhibit a range in SFR$_{UV,corr}$ between 0.2 $\Msun$yr$^{-1}$ and $\sim$2000 $\Msun$yr$^{-1}$. A full description of our star formation measure is provided in Bauer et al. (2011).

\section{The stellar mass functions}
\subsection{Total Stellar Mass Function}
\label{sec:totalMF}
\begin{figure*}
\centering
\includegraphics[trim = 4mm 0mm 0mm 3mm, clip,scale=0.85]{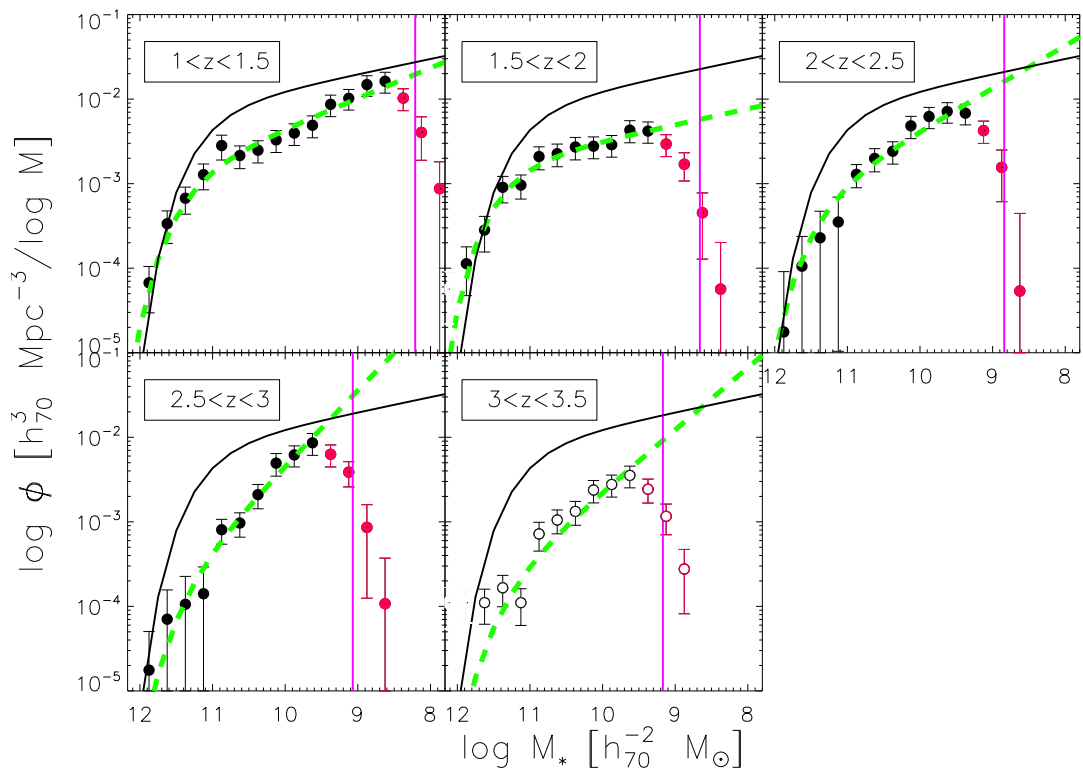}
\caption{The galaxy stellar mass function between redshifts $z=1-3.5$ in the GNS. The dashed green curve is the Schechter fit to the data. The black points are the data that has been fitted, and the red points are the data that has been left out of the fit due to incompleteness. The final redshift bin is represented by open circles as it is not included in the bulk of our analysis as discussed in Section \ref{sec:totalMF}. The solid pink vertical lines show the theoretical mass limits of the GNS survey (see Section \ref{sec:biascomp}). Also included for comparion is the solid black line which represents the local galaxy stellar mass function of \citet{Cole01}. The data used to construct these mass functions can be found in Table \ref{tab:gsmfdata}}
\label{totMF}
\end{figure*}

\begin{figure*}
\centering
\includegraphics[trim = 15mm 55mm 0mm 3mm, clip,scale=1.1]{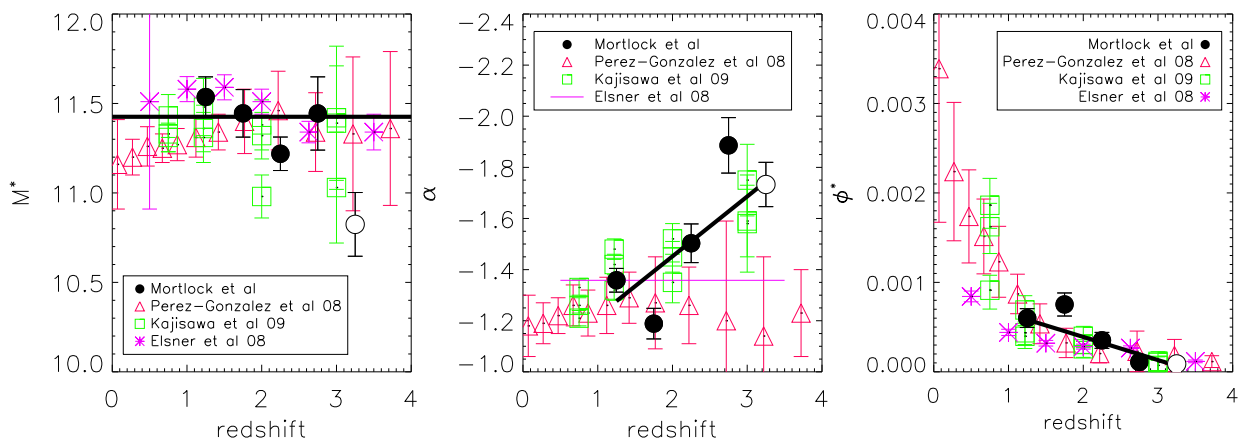}
\caption{The parameter results of the Schechter fits. Left: The results of the fitting for $M^{*}$ with all parameters free, Middle: The result of the fitting for $\alpha$ with $M^{*}$ held constant (see Table \ref{tab:1}), and Right: The results of the fitting for $\phi^{*}$ with $M^{*}$ held constant. For each panel the black circles are the results from this work, the red triangles are \citet{Pere08} results, the green squares are the \citet{Kaji09} results (from three different SED models) and the pink stars (pink line for middle panel) are the \citet{Elsn08} results. For each panel the final redshift point is plotted as an open circle as it is not considered in the analysis. The  parameters of the final redshift bin are represented by open circles as they are not included in the bulk of our analysis as discussed in Section \ref{sec:totalMF}.} 
\label{totparams}
\end{figure*}

\begin{table}
\begin{tabular}{|r|r|r|}
\hline
  \multicolumn{1}{|c|}{Redshift Range} &
  \multicolumn{1}{c|}{log $M_{*} [h_{70}^{-2} M_{\odot}]$} &
  \multicolumn{1}{c|}{log $\phi [h_{70}^{3} Mpc^{-3}/log M]$} \\
\hline
 1.0 to 1.5 & 8.6 & -1.79 $\pm$ 0.12\\
 & 8.9  & -1.83 $\pm$ 0.12\\
 & 9.1  & -1.99 $\pm$ 0.11\\
 & 9.4  & -2.06 $\pm$ 0.12\\
 & 9.6  & -2.31 $\pm$ 0.13\\
 & 9.9  & -2.40 $\pm$ 0.12\\
 & 10.1 & -2.48 $\pm$ 0.13\\
 & 10.4 & -2.60 $\pm$ 0.13\\
 & 10.6 & -2.67 $\pm$ 0.13\\
 & 10.9 & -2.55 $\pm$ 0.14\\
 & 11.1 & -2.89 $\pm$ 0.15\\
 & 11.4 & -3.17 $\pm$ 0.15\\
 & 11.6 & -3.47 $\pm$ 0.18\\
 & 11.9 & -4.17 $\pm$ 0.24\\
 1.5 to 2.0 & 9.4 & -2.38 $\pm$ 0.12\\
 & 9.6  & -2.37 $\pm$ 0.13\\
 & 9.9  & -2.54 $\pm$ 0.12\\
 & 10.1 & -2.56 $\pm$ 0.13\\
 & 10.4 & -2.57 $\pm$ 0.13\\
 & 10.6 & -2.64 $\pm$ 0.13\\
 & 10.9 & -2.68 $\pm$ 0.13\\
 & 11.1 & -3.02 $\pm$ 0.14\\
 & 11.4 & -3.04 $\pm$ 0.15\\
 & 11.6 & -3.55 $\pm$ 0.20\\
 & 11.9 & -3.95 $\pm$ 0.25\\
 & 12.1 & -4.25 $\pm$ 0.27\\
  2.0 to 2.5 & 9.4 & -2.17 $\pm$ 0.12\\
 & 9.6  & -2.15 $\pm$ 0.12\\
 & 9.9  & -2.21 $\pm$ 0.12\\
 & 10.1 & -2.32 $\pm$ 0.13\\
 & 10.4 & -2.62 $\pm$ 0.13\\
 & 10.6 & -2.70 $\pm$ 0.13\\
 & 10.9 & -2.89 $\pm$ 0.13\\
 & 11.1 & -3.45 $\pm$ 0.42\\
 & 11.4 & -3.64 $\pm$ 0.46\\
 & 11.6 & -3.98 $\pm$ 0.54\\
 & 11.9 & -4.75 $\pm$ 1.81\\
  2.5 to 3.0 & 9.6 & -2.07 $\pm$ 0.13\\
 & 9.9  & -2.21 $\pm$ 0.12\\
 & 10.1 & -2.31 $\pm$ 0.13\\
 & 10.4 & -2.68 $\pm$ 0.14\\
 & 10.6 & -3.01 $\pm$ 0.14\\
 & 10.9 & -3.09 $\pm$ 0.14\\
 & 11.1 & -3.85 $\pm$ 0.47\\
 & 11.4 & -3.98 $\pm$ 0.49\\
 & 11.6 & -4.15 $\pm$ 0.53\\
 & 11.9 & -4.75 $\pm$ 0.81\\
  3.0 to 3.5 & 9.6 & -2.45 $\pm$ 0.12\\
 & 9.9  & -2.56 $\pm$ 0.13\\
 & 10.1 & -2.62 $\pm$ 0.13\\
 & 10.4 & -2.88 $\pm$ 0.14\\
 & 10.6 & -2.98 $\pm$ 0.14\\
 & 10.9 & -3.14 $\pm$ 0.16\\
 & 11.1 & -3.96 $\pm$ 0.20\\
 & 11.4 & -3.78 $\pm$ 0.17\\
 & 11.6 & -3.96 $\pm$ 0.19\\
 & 11.9 & -3.96 $\pm$ 0.19\\
 & 12.1 & -3.96 $\pm$ 0.22\\
\hline\end{tabular}
\centering
\caption{The total galaxy stellar mass functions seen in Figure \ref{totMF}}
\label{tab:gsmfdata}
\end{table}

\begin{table*}
\begin{tabular}{ | c | c | c | c | c | c | c | }
\hline
  Redshift Range & $M^{*}$ & \(\phi\)$(\times10^{-4})$ & \(\alpha\)  \\
\hline
  1.0 to 1.5 & 11.43 & 6.01  $\pm$ 1.05 & -1.36 $\pm$ 0.05 \\
  1.5 to 2.0 & 11.43 & 7.53  $\pm$ 1.23 & -1.19 $\pm$ 0.06 \\
  2.0 to 2.5 & 11.43 & 3.52  $\pm$ 0.89 & -1.50 $\pm$ 0.08 \\
  2.5 to 3.0 & 11.43 & 1.11  $\pm$ 0.36 & -1.89 $\pm$ 0.11 \\
  3.0 to 3.5 & 11.43 & 0.89  $\pm$ 0.22 & -1.73 $\pm$ 0.09 \\ 

\hline
\end{tabular}
\centering
\caption{The values of the parameters from the Schechter fit. $M^{*}$ is the mean value from fitting with all parameters free. $\phi^{*}$ and $\alpha$ are the result of the Schechter fit with $M^{*}$ held constant.}
\label{tab:1}
\end{table*}

To construct galaxy stellar mass functions our data has been split into five redshift bins between $z=1$ and $z=3.5$. These are chosen to have roughly the same comoving volume to reduce fluctuations in the total number of galaxies. The highest redshift bin, $z=3$ to $3.5$, contains around 300 galaxies, thus we maintain a good sample size throughout. However, we exclude the final redshift bin from much of our analysis as, in that redshift range, the H-band is no longer sampling primarily optical lightand no longer provides a good measure of the balmer break, thus we find larger errors in the measurement of stellar masses.

We calculate the number densities, i.e. the number of galaxies per co-moving volume per mass interval, and plot the stellar mass functions for each redshift bin in Figure \ref{totMF}. Plotted for comparison in Figure \ref{totMF}, as a solid black curve, is the local stellar mass function from \citet{Cole01}. Also plotted is the vertical line that indicates the mass limit of the survey in each redshift bin (explanation of the calculation for this can be found in Section \ref{sec:biascomp}). These lower limits are at a very low stellar mass compared with other work such as \citet{Marc09}, \citet{Pere08} and \citet{Font06}, who do not probe stellar masses as low as we do (see Section \ref{sec:diss} for discussion of this).

\subsubsection{High Mass Bias and Completeness}
\label{sec:biascomp}
The GNS is specifically intended to maximise the population of high mass galaxies, and hence the galaxy sample will contain more galaxies with $M_{*}>10^{11} M_{\odot}$ than expected for a randomly positioned survey. The high mass end of the stellar mass function is corrected for this by computing the ratio of high  mass galaxies in the GNS pointings to the number of high mass galaxies in the total GOODS fields. By comparing this to the fraction of the area that the pointings covered, we obtain a correction factor for the overdensity of galaxies with $M_{*}>10^{11} M_{\odot}$, finding a value of 3.05. The high stellar mass galaxies ($M_{*}>10^{11} M_{\odot}$) in the redshift bins 2.0 to 2.5 and 2.5 to 3.0 were thus divided by 3.05 before fitting (\citealt{Cons10}).

Another problem in constructing the stellar mass functions is incompleteness at the low mass end. As mentioned previously, Figure \ref{totMF} shows the mass limit of the survey. This is a purely theoretical mass limit calculated from the central wavelength of the NICMOS camera. We use this to calculate the rest frame band observed by NICMOS for each redshift bin. From this we calculate the mass to light ratios of a maximally old stellar population, again for each redshift range. We combined this with the luminosity limit calculated from the limiting magnitude of the survey and the luminosity distance for each redshift bin. This gave us the mass limit of the survey represented by the pink vertical line in Figure \ref{totMF}.

Figure \ref{totMF} shows that galaxies appear to drop out before the calculated mass limit. This is due to the pipeline detection being less sensitive to galaxies at the low mass end (\citealt{Cons10}), thus we do not include these points in our fits. To determine when this occurs we look at the residuals between the local mass function and the calculated number densities. A change in trend of the residual represents the loss of completeness and we do not fit below this part of the stellar mass function. The points not included in the fit are the red points in Figure \ref{totMF}

\subsubsection{Fitting the Schechter Function}
\label{sec:fitting}
We fit our stellar mass functions, within the errors on the number densities, with a Schechter function (\citealt{Sche76}) of the form

\begin{equation}
\phi(M) = \phi^{*}\cdot \rm log(10)\cdot[10^{(M-M^{*})}]^{(1+\alpha)}\cdot \rm exp[-10^{(M-M^{*})}].
\label{eq:sch}
\end{equation}
In this equation the parameter $M^{*}$ is the characteristic mass at which the stellar mass function turns over, $\alpha$ parameterises the slope of the faint end of the stellar mass function, and $\phi^{*}$ is the scale factor. The first fitting of the stellar mass functions we performed was done by leaving all of the three parameters free. For the redshift bins $z=2-2.5$ and $z=2.5-3$ we find that the highest mass points ($M_{*}>10^{11} M_{\odot}$) have extremely large errors. To help constrain the high mass end we combine the highmass points into one point in these redshift bins.

The left hand panel of Figure \ref{totparams} shows the results of $M^{*}$ for the first fitting, where we find there is very little change in $M^{*}$ over redshift (except at $z=3-3.5$). This is in good agreement with various other studies such as \citet{Pere08} and \citet{Elsn08} (these points can also be seen in the left hand panel of Figure \ref{totparams}). We then repeat the fitting holding $M^{*}$ constant at its mean value (see Table \ref{tab:1}) over the whole redshift range, which allows for a better constraint on $\alpha$ and $\phi^{*}$. Some previous work have found a constant $\alpha$ up to $z\sim 2$ (e.g. \citealt{Font04} and \citealt{Borc06}). In this work we find no evidence of this, hence we only use a constant $M^{*}$ for investigating the evolution of $\alpha$ and $\phi^{*}$ at $1<z<3$. The values of the fitting parameters are shown in Table \ref{tab:1} and also compared to previous work in Figure \ref{totparams}. For this repeated fitting we do not combine the high mass points as before, as the high mass end of the stellar mass function is already well constrained by $M^{*}$.

\subsubsection{Errors and Simulations}
\label{sec:err}
We calculate errors on our number densities using two Monte Carlo simulations taking into account the 1$\sigma$ Gaussian measured error on the stellar masses and accounting for errors on the redshifts. We first use the measured error on each stellar mass and compute a Gaussian distribution of simulated stellar masses, for each galaxy, between $\pm3\sigma$ of their measured error. Then a new stellar mass was randomly selected from this Gaussian distribution, so that we obtain a simulated stellar mass for each galaxy. We then recalculate the number densities in the same way as we did for the original catalogue of stellar masses, and the error due to the catalogue mass error, is the difference between the simulated number density and the original catalogue number density.

For the photometric redshift errors we construct a catalogue of simulated redshifts using the same Gaussian method we use to obtain the simulated stellar masses described previously. We assume Gaussian errors on the photometric redshifts, and do not consider the catastrophic outliers. We then calculate the luminosity distance for both the original and the simulated redshifts of each galaxy. We calculate the ratio of these two luminosity distances and assume this is equal to the ratio of stellar masses. By multiplying the original galaxy stellar mass by the luminosity distance squared ratio, we obtain a new simulated galaxy stellar mass. As before we then recalculated the number densities using this catalogue of simulated stellar masses, then subtract the original stellar mass from the simulated stellar mass to obtain the error. We then add the stellar mass and redshift errors in quadrature, as well as the error due to Poisson statistics, to obtain the total error on each number density. These are the errors shown on Figure \ref{totMF}.

We also simulate independently the effects of the measured errors on the stellar mass function by randomly altering each stellar mass by $\pm$0.25 dex. We chose this number to investigate the extremes of the effect of large errors. This shows how a mass bin containing less galaxies will be effected by the errors on the stellar mass more than a fuller mass bin. This is due to the Eddington bias, i.e. if a mass bin is relatively empty, galaxies from nearby fuller bins will spill over in to these bins more readily due to measurement errors. We reanalyse the altered masses and find that the largest variation is in the emptier, high mass bins. These variations lie within the calculated errors on the number densities and thus we can still reconstruct the same Schechter function parameters.

\subsubsection{Inspection of the Mass Functions}
\label{sec:inspec}
There are several features that can be noted through inspection of the stellar mass functions, and from the best fit parameters. Firstly, the massive galaxies ($M_{*}>10^{11} M_{\odot}$) are present, with a similar number density as at $z=0$, up to a redshift $z=3$. The low mass galaxies do not reach the local value of the number density until after the massive galaxies.  This is a downsizing in terms of stellar mass over a large range. We discuss this further in Section \ref{sec:diss}.

The intermediate stellar mass galaxies ($M_{*} \sim 10^{9.5} M_{\odot}$ to $M_{*} \sim 10^{11} M_{\odot}$) show a decreased rate of formation compared to the lower stellar mass galaxies. This manifests itself most clearly in the lowest redshift bin as a dip in the galaxy stellar mass function in the intermediate stellar mass range. This feature is also possibly present between redshifts of $z=2-3$, although less obvious. We discuss this feature in Section \ref{sec:diss}

As noted before we find a general trend for $\alpha$ is to increase at higher redshift, as shown by the black line in the middle panel of Figure \ref{totparams}. We also find that our values of $\alpha$ are more negative (therefore steeper) than found in previous studies. The middel panel of Figure \ref{totparams} shows that our results for  $\alpha$ are steeper than \cite{Elsn08} and \cite{Pere08}. This is also the case for other studies, including \cite{Font04} (held $\alpha$ constant at $-1.27/-1.36$ from $z=1-2$) and \cite{Font06} ($\alpha=-1.27$ to $-1.47$ from $z=1.15-3.5$). We do however find a similarity between our results and \cite{Kaji09} who explore a similar redshift range and stellar mass depth to us. We find that $\phi^{*}$ is decreasing at higher redshift, as demonstrated by the straight line fit to the right panel of Figure \ref{totparams}. The overall decrease in $\phi^{*}$ represents the overall decrease in number density, as is expected at higher redshifts since fewer galaxies have had time to form.

\begin{figure*}
\centering
\includegraphics[trim = 4mm 0mm 0mm 3mm, clip,scale=0.85]{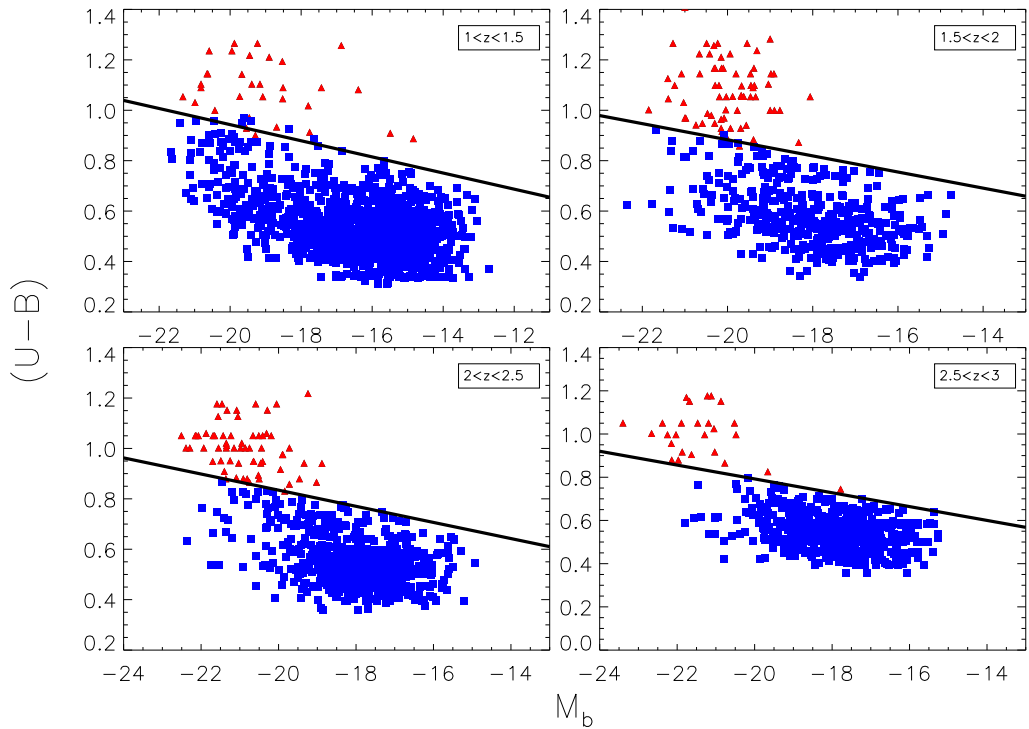}
\caption{The division of the red and blue galaxy populations. The solid black line represents the colour cut used to divide the galaxies into their respective populations corrected for redshift evolution.}
\label{UBvsMb}
\end{figure*}

\begin{figure*}
\centering
\includegraphics[trim = 4mm 0mm 0mm 3mm, clip,scale=0.85]{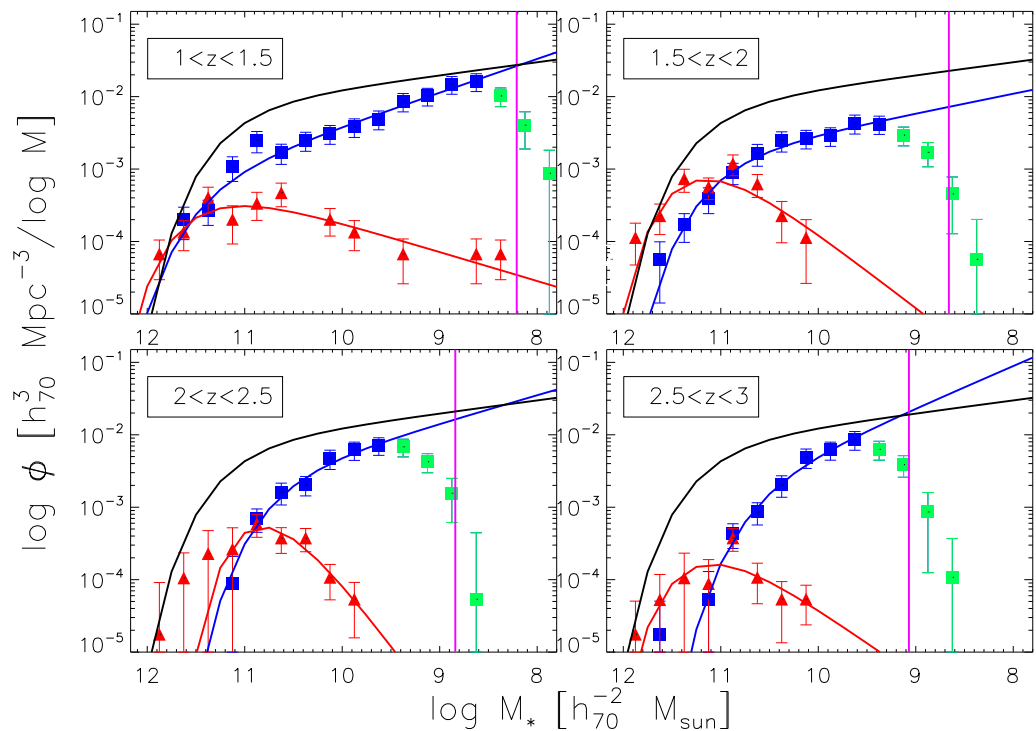}
\caption{The colour divided galaxy stellar mass functions. The red and blue points represent stellar mass functions of the red and blue galaxy populations as defined in Section \ref{sec:MFforBR}. The green points are the points not included in the Schechter fit. The solid red and blue curves are the Schechter fits. The solid black curve is the local galaxy stellar mass function of \citet{Cole01}. The solid pink vertical line is the theoretical mass limit of this survey (see Section \ref{sec:biascomp}).}
\label{BRMF}
\end{figure*}

\begin{figure*}
\centering
\includegraphics[trim = 15mm 55mm 0mm 3mm, clip,scale=1.1]{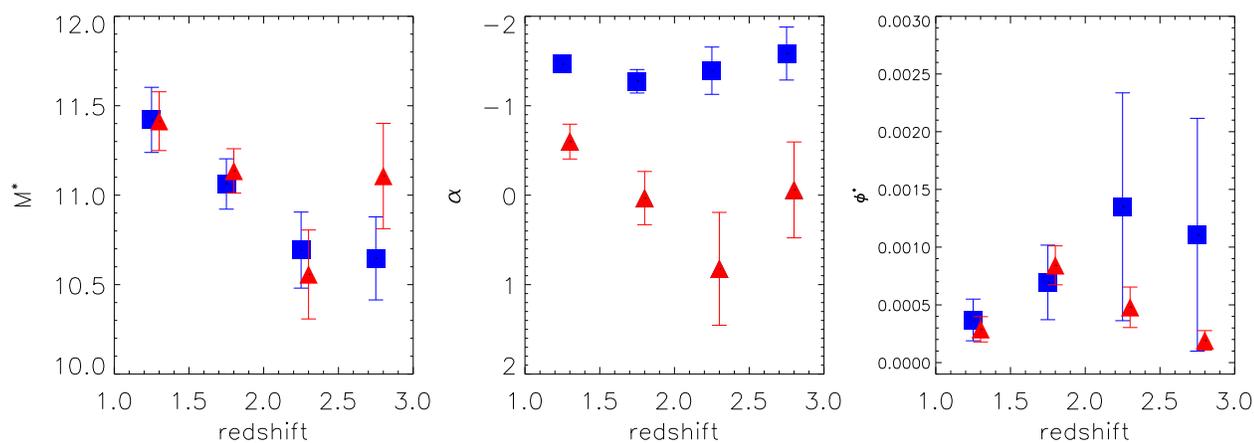}
\caption{The Schechter fit parameters of the blue and red galaxy stellar mass functions. Left: The results of the fitting for $M^{*}$. The blue/red points are $M^{*}$ for the blue/red galaxy populations when all of the parameters are free. Middle: The result of the fitting for $\alpha$. The blue/red points are $\alpha$ for the blue/red galaxy populations when $M^{*}$ is held constant. Right: The result of the fitting for $\phi^{*}$. The blue/red points are $\alpha$ for the blue/red galaxy populations when $M^{*}$ is held constant. All the red points are offset by 0.05 in redshift for clarity.}
\label{BRpars}
\end{figure*}

\begin{table*}
\begin{tabular}{ | c | c | c | c | c | c | c | c |}
\hline
  Redshift Range &  & $M^{*}$ & \(\phi\)$(\times10^{-4})$ &  \(\alpha\)  \\
\hline
  1.0 to 1.5 & Blue & 11.51 $\pm$ 0.15 & 3.07 $\pm$ 1.35 & -1.48 $\pm$ 0.06\\
             & Red  & 11.41 $\pm$ 0.16 & 2.88 $\pm$ 1.09 & -0.60 $\pm$ 0.19\\
  1.5 to 2.0 & Blue & 11.06 $\pm$ 0.14 & 6.94 $\pm$ 3.23 & -1.27 $\pm$ 0.13\\
             & Red  & 11.13 $\pm$ 0.12 & 8.42 $\pm$ 1.69 &  0.03 $\pm$ 0.30\\
  2.0 to 2.5 & Blue & 10.68 $\pm$ 0.21 & 14.19$\pm$ 10.16& -1.38 $\pm$ 0.26\\
             & Red  & 10.58 $\pm$ 0.24 & 4.79 $\pm$ 1.75 &  0.82 $\pm$ 0.63\\
  2.5 to 3.0 & Blue & 10.67 $\pm$ 0.24 & 10.21$\pm$ 9.40 & -1.60 $\pm$ 0.29\\
             & Red  & 11.11 $\pm$ 0.29 & 1.92 $\pm$ 0.85 & -0.06 $\pm$ 0.54\\
\hline
\end{tabular}
\centering
\caption{The values of the parameters from the Schechter fit to the blue and red galaxy population.}
\label{tab:2}
\end{table*}

\subsection{Blue/Red Mass Functions}
\label{sec:MFforBR}
To obtain the galaxy stellar mass functions for the blue and red galaxy populations we divided the galaxies by colour. This was done by dividing the sample using the red sequence equation,
\begin{equation}
(U-B) = -0.032(M_{B} + 21.52) + 1.284 - 0.25,
\label{eq:colcut}
\end{equation}
from \citet{Will06}, modified for the AB magnitude system. The equation applies at redshift of $z=1.0$, hence we modified Equation \ref{eq:colcut} to account for redshift evolution. This was done by using the redshift evolution of the luminosities and colours of galaxies from \citet{Vand01}. This allowed us to obtain a change in $M_{B}$ and $(U-B)$ at higher redshifts due to passive evolution. The change is then included in the equation so that the right hand side of Equation \ref{eq:colcut} becomes 
\begin{equation}
-0.032(M_{B}-\Delta M_{B}+21.52)+1.284-0.25+\Delta(U-B).
\label{eq:colcut2}
\end{equation}
We then apply the cut so that if $(U-B)$ is greater than Equation \ref{eq:colcut2} the galaxy is red, and if $(U-B)$ is less than Equation \ref{eq:colcut2} the galaxy is blue. The results can be seen in Figure \ref{UBvsMb}, where the red/blue triangles/squares are the red/blue galaxies and the black line is the colour cut.

After applying this colour cut we split the blue and red galaxy populations into the same redshift bins as before, and the number densities computed in the same way for each colour. Also the errors on these mass functions are computed exactly as before using our Monte Carlo approach (Section \ref{sec:err}). The resulting colour stellar mass functions are shown in Figure \ref{BRMF}. Here the blue and red solid lines represent the Schechter fits to the blue and red galaxy populations, and the solid black curve is the local galaxy stellar mass function of \citet{Cole01}. The green points are the points that are not included in the fitting of the blue population. Both the blue and red stellar mass functions are fit using equation \ref{eq:sch}. In this case we see no evidence for a constant $M^{*}$, and thus we leave all parameters free when fitting. This is discussed further in Section \ref{sec:diss} in terms of implied evolution.

\subsubsection{Inspection of the Blue/Red Mass Functions}
The colour stellar mass functions in Figure \ref{BRMF}, like the total stellar mass functions, show stellar mass downsizing. We see that even in the redshift range $z=2.5-3$, red and blue galaxies at $M_{*}>10^{11.5} M_{\odot} $ are present with number densities very close to the local value. In the redshift bin $z=1-1.5$ we see that both the red and the blue galaxies with $M_{*}>10^{11.5} M_{\odot} $ are nearly fully in place, whereas the low stellar mass galaxies have not formed as quickly. 

The low stellar mass end shown in Figure \ref{BRMF} is dominated by blue galaxies at $z<3$, and their number densities are close to the local stellar mass function out to a redshift of $z=3$. This means that the steepness of the slope in the total stellar mass function (Figure \ref{totMF}) is dominated by low mass blue galaxies. For the blue galaxies themselves $\alpha$ remains roughly constant and $\phi^{*}$ is constant within the error bars. This is shown in  Figure \ref{BRpars}. The parameter $M^{*}$ on the other hand, shows a general decline unlike what we see in the total stellar mass functions.

For the red population, the fitted value of $\alpha$ shows a general increase (less steep) at higher redshifts, which is also the case for the parameter $M^{*}$. Contrary to this, the parameter $\phi^{*}$ shows very little variation. Unfortunately, we do not have good number statistics for the stellar mass functions of the red population (in the range $z=2.5-3$ there are 26 red galaxies compared to 639 blue). We also have large errors on the high mass galaxies over $z=2-3$, and hence the fit to the red stellar mass functions do not provide robust results. To this end, we cannot make any strong conclusions about the evolution of the red galaxy population.

\subsection{The Mass Functions of Star Forming and Non Star Forming Galaxies}
\label{sec:MFforSF}
Having previously examined the colours of the GNS galaxies we next investigate the differences or similarities between colour and star formation selected stellar mass functions. We use the star formation rates calculated as described in Section \ref{sec:SFRexplan} for the sample between $z=1.5-3$ and for $M_{*}>10^{9.5} M_{\odot} $. To divide the galaxies into passive and evolving populations we use the star formation rate divided by stellar mass of the galaxy to calculate the time it would take for a galaxy to double in size, we call this $t_{\rm double}$. Using the Hubble time ($t_{h}$) at the redshift of each galaxy we obtain a measure of how fast a galaxy is forming by calculating $t_{\rm form}$, where
\begin{equation}
t_{\rm form} = t_{\rm double}/t_{h}.
\label{eq:tform}
\end{equation}
We averaged over all the values of $t_{\rm form}$ to cut our sample into two distinct populations. Those above an average of $\langle t_{\rm form}\rangle $=0.1 are considered non-star forming, those below are considered star forming. The resulting stellar mass functions can be seen in Figure \ref{DERCUT}.

The general trend in Figure \ref{DERCUT} shows an increase of both populations at high stellar mass. In this region $M_{*}>10^{11} M_{\odot}$ we find the non-star forming galaxies dominate. At the low stellar mass end we see that the star forming population dominates over the non-star forming galaxies. The slope of the low stellar mass end is very steep with this steepness decreasing over time, even within the large error bars.

\subsubsection{Comparison Between the Blue/Red and the Star Forming/Non Star Forming Mass Functions}
\begin{figure*}
\centering
\includegraphics[trim = 16mm 55mm 0mm 3mm, clip,scale=1.2]{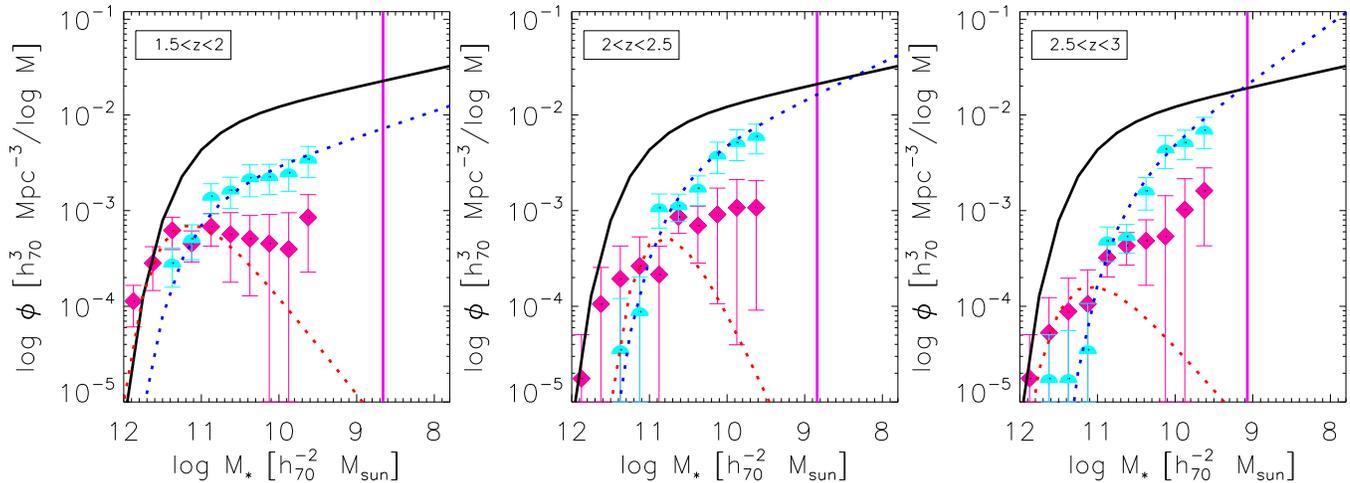}
\caption{The galaxy stellar mass functions of star forming and non-star forming populations, defined by Equation (\ref{eq:tform}). The pink diamonds show the non-star forming galaxy population and the blue half circles are star forming galaxies. The solid black curve is the local galaxy stellar mass function of \citet{Cole01}. The solid pink vertical line is the mass limit of this survey. The red and blue dashed lines are the Schechter fits for the red and blue galaxy populations plotted for comparison (see also Figure \ref{BRMF}).}
\label{DERCUT}
\end{figure*}

Over plotted on Figure \ref{DERCUT} are the Schechter fits for the blue and red galaxy populations, shown as the blue and red dotted lines. We find that over all the redshift ranges the star forming/non-star forming galaxies match the blue/red populations within the errors plotted. However there is some slight disagreement at the high mass end of the mass functions in the range $z=2-3$. The highest mass star forming galaxies are better represented by the red Schechter function, this is likely the result of the presence of dusty star formation (Bauer et al. (2011), \citet{Grut10}.

\section{Integrated Stellar Mass Densities}
\label{sec:massden}
\begin{figure}
\centering
\includegraphics[trim = 8mm 0mm 0mm 3mm, clip,scale=0.65]{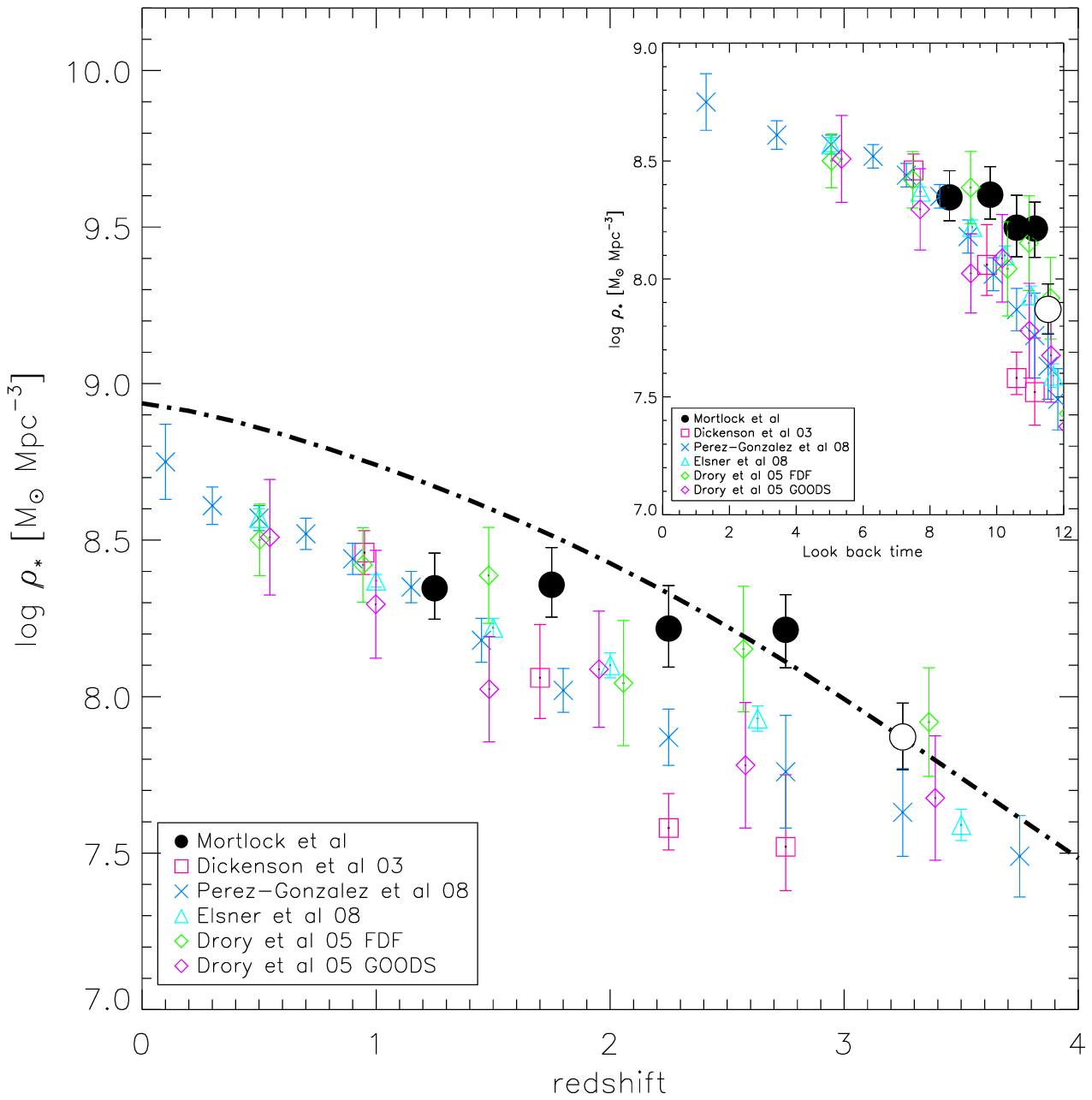}
\caption{The integrated stellar mass density calculated for each redshift bin using the integration of the Schechter function. The black circles show the results of this work. The pink squares are from \citet{Dick03}, the dark blue crosses are from \citet{Pere08}, the light blue triangles are from \citet{Elsn08} and the green and purple diamonds are from \citet{Dror05}.The black dashed dot line is a prediction of the stellar mass densities from the integrated star formation rate (\citealt{Wilk08}). The final redshift point is plotted as an open circle as it is not considered in the analysis as discussed in Section \ref{sec:totalMF}.}
\label{SMD}
\end{figure}

\begin{figure}
\centering
\includegraphics[trim = 8mm 0mm 0mm 3mm, clip,scale=0.65]{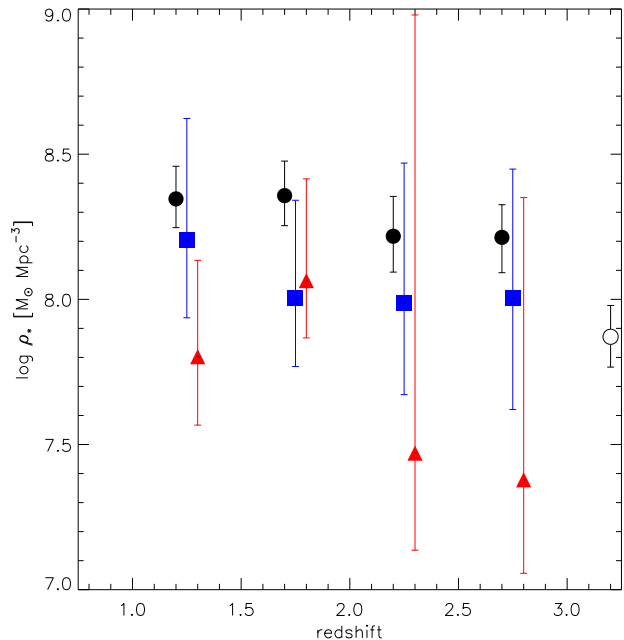}
\caption{The integrated stellar mass density calculated for each redshift bin using the integration of the Schechter function for the red and blue galaxy populations. The black circles are the total stellar mass density plotted for reference. The black points are offset in redshift by -0.05 and the red points are offset in redshift by +0.05. The final redshift point for the total population is plotted as an open circle as it is not considered in the analysis as discussed in Section \ref{sec:totalMF}.}
\label{SMDbr}
\end{figure}

\begin{table*}
\begin{tabular}{ | c | c | c | c | }
\hline
\vspace{3 mm}
&\multicolumn{3}{|c|}{log $\rho_{*} [M_{\odot} Mpc^{-3}]$} \\
  Redshift Range & $M_{*}=10^{7} M_{\odot}$ Limit & $M_{*}=10^{9.5} M_{\odot}$ Limit & $M_{*}=10^{3} M_{\odot}$ Limit\\
\hline
  1.0 to 1.5 & 8.35$^{+0.11}_{-0.10}$ & 8.32 $^{+0.12}_{-0.11}$ & 8.35 $^{+0.11}_{-0.10}$\\
  1.5 to 2.0 & 8.36$^{+0.12}_{-0.10}$ & 8.34 $^{+0.12}_{-0.11}$ & 8.36 $^{+0.12}_{-0.10}$\\
  2.0 to 2.5 & 8.23$^{+0.14}_{-0.12}$ & 8.16 $^{+0.16}_{-0.14}$ & 8.22 $^{+0.14}_{-0.12}$\\
  2.5 to 3.0 & 8.21$^{+0.11}_{-0.12}$ & 7.95 $^{+0.18}_{-0.16}$ & 8.33 $^{+0.06}_{-0.08}$\\
  3.0 to 3.5 & 7.87$^{+0.11}_{-0.10}$ & 7.72 $^{+0.15}_{-0.13}$ & 7.90 $^{+0.09}_{-0.09}$\\ 
\hline
\end{tabular}
\centering
\caption{The values of the parameters from the Schechter fit. $M^{*}$ is the mean value from fitting with all parameters free. $\phi^{*}$ and $\alpha$ are the result of the Schechter fit with $M^{*}$ held constant.}
\label{tab:3}
\end{table*}

We integrate over the Schechter function to get the integrated stellar mass density between $z=1-3$. This is the total amount of stellar mass contained within galaxies, in a given redshift range, per comoving volume. The results of this calculation are shown in Figure \ref{SMD}. To calculate this we perform a numerical integration between the limits of $M_{*}=10^{12} M_{\odot}$ and $M_{*}=10^{7} M_{\odot}$. We extended the integration beyond the lower mass limit of our survey by extrapolating the Schechter function to masses beyond those which we fit.

Our stellar mass densities are generally higher than what has been found in previous work, as shown in Figure \ref{SMD}, due to our steeper values of $\alpha$. The black dashed dot line shows the stellar mass density history obtained from the integration of the instantaneous cosmic star formation history computed by \citet{Wilk08}, using the same IMF as in our work. They show that at $z>0.7$ the line does not agree well with previous work which inferred from the integration of the stellar mass at these redshifts. The stellar mass densities that we calculate are higher than what has been found in previous works, but are still slightly systematically lower than the integrated star formation history line.

We also computed the stellar mass density with a bright lower limit of $M_{*}=10^{9.5} M_{\odot}$ and a faint lower limit of $M_{*}=10^{3} M_{\odot}$. The bright limit allows us to compare our computed stellar mass densities with the mass we can actually observe in all mass bins, and the faint mass limit allows a better comparison with the integrated cosmic star formation history. We find that when we adopt these bright and faint limits we see very little difference between these the computed stellar mass densities. Even in the range $z=2.5-3$, where the largest difference is found, the stellar mass density from both the bright and faint limit still lie within the errors of the original stellar mass density.

The stellar mass densities found in this work can be compared to the local value of \citet{Cole01}. We find therefore that by $z\sim 1.25$, 40$^{+10}_{-9}$\% of the stellar mass in the local universe has formed and by $z\sim 1.75$, 41$^{+11}_{-10}$ \% has formed. The growth of stellar mass does not alter dramatically after $z=2$. At $z= 2.25$, 30$^{+10}_{-9}$\% has formed, and we see that at $z= 2.75$, 30$\pm8$\% is in place. The final stellar mass density at $z= 3.25$ is 14 $\pm3$\% of the local value, although the results for this redshift range are not as robust as the others, as discussed in Section \ref{sec:totalMF}. Thus, for $1<z<3$, we see that the stellar mass density changes only slightly, hence there is very little change in stellar mass over this time period. We also note the stellar mass forms quickly as roughly one third of the stellar mass is already formed when the Universe is only 2.3 Gyrs old.

Figure \ref{SMDbr} shows the stellar mass density for the blue and red galaxy populations, with the total stellar mass density shown in black. We see a large decrease in the mass density for the red galaxies, yet for the blue galaxies the stellar mass density is consistent with being constant over this redshift range. This is seen in other studies such as \citet{Borc06}, \citet{Arno07} and \citet{Ilbe09} who combined find very little evolution in the blue stellar mass functions as far out as redshift of $z\sim 2$. For the red population the numbers themselves are not as robust due to the poor fitting, but the different evolution of the red galaxies is clear. This growth in stellar mass density in the red population is consistent with other works such as \citet{Arno07} and \citet{Ilbe09}.
\begin{figure*}
\centering
\includegraphics[trim = 4mm 0mm 0mm 3mm, clip,scale=0.85]{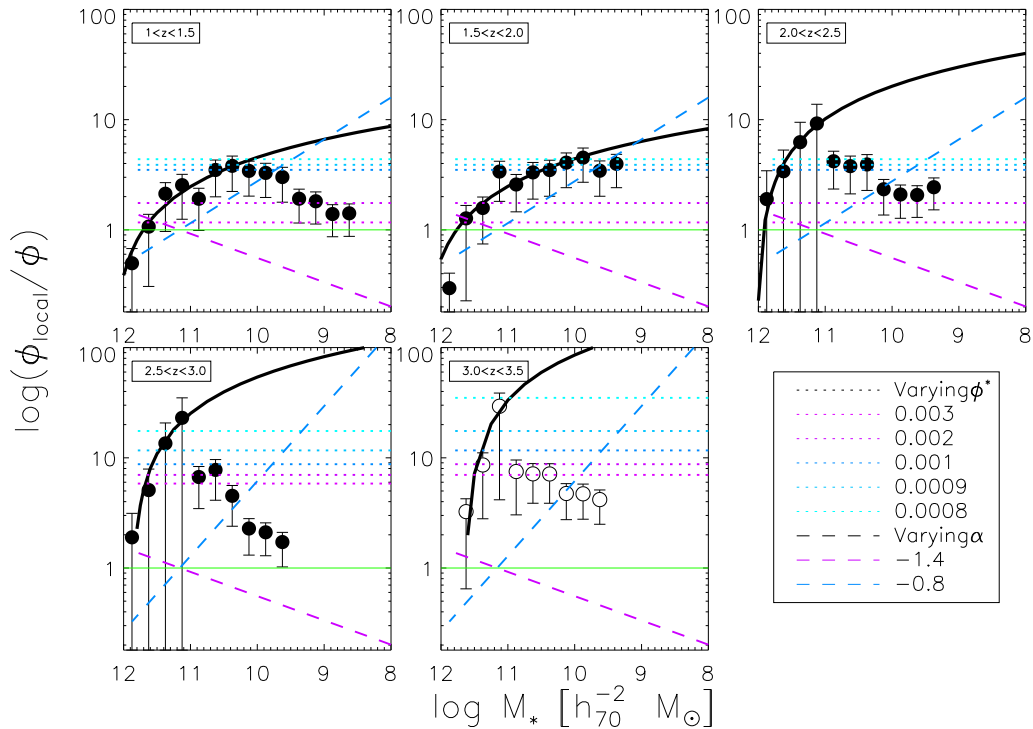}
\caption{The ratio between the local galaxy stellar mass functions of \citet{Cole01} and the number densities of the total galaxies in this work. The green line is the local stellar mass function over the local stellar mass function. The dotted line shows the local stellar mass function with varying $\phi^{*}$ and the dashed line shows the local stellar mass function with varying $\alpha$. The final redshift bin is represented by open circles as it is not included in the bulk of our analysis as discussed in Section \ref{sec:totalMF}.}
\label{totresid}
\end{figure*}

\begin{figure*}
\centering
\includegraphics[trim = 4mm 0mm 0mm 3mm, clip,scale=0.85]{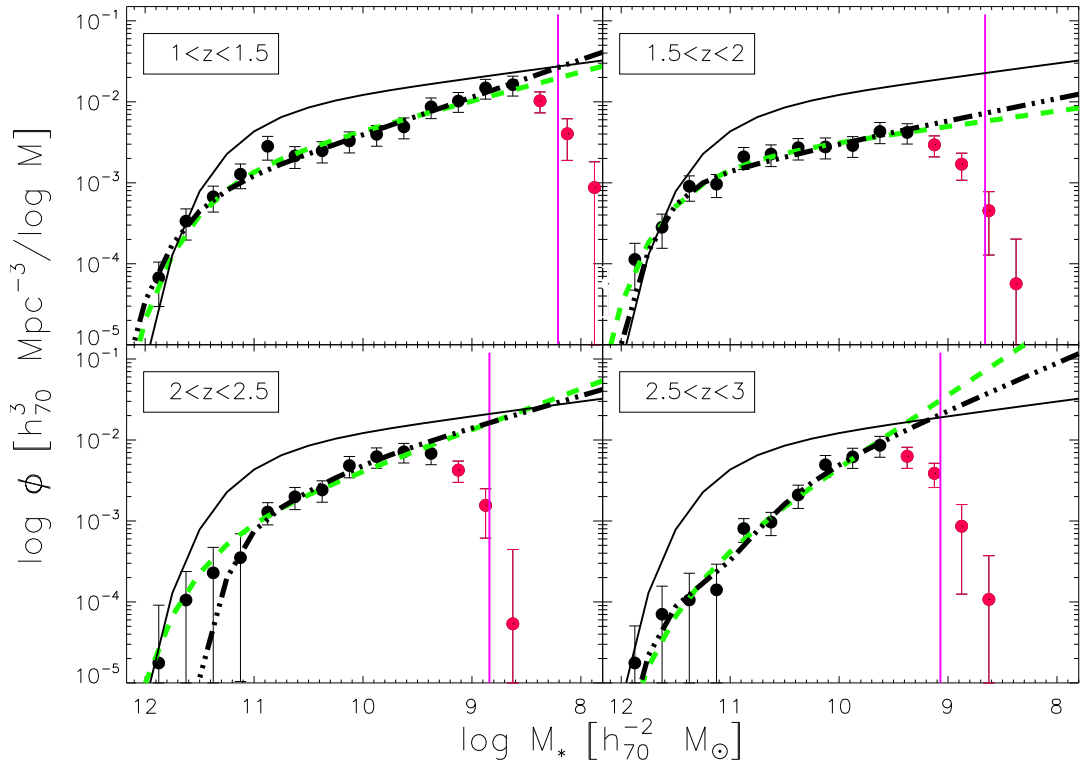}
\caption{The total Schechter functions (green dashed line). Over plotted is a double Schechter function (black dash dotted line) calculated using the blue and red fit added together.}
\label{totcomp}
\end{figure*}

\section{Discussion}
\label{sec:diss}
\subsection{Comparison With Other Surveys and Data}
There is a large amount of research exploring the stellar mass function over a large range in redshifts, from $z=0-6$. What is unique about the GNS, and this work in particular, is the depth and redshift range provided by the data. A large number of surveys (e.g. COSMOS, zCOSMOS and COMBO-17) previously examine the stellar mass function at a lower redshift range than the GNS. For $z<1$ these surveys probe a large stellar mass range e.g. \citet{Pozz09}, with mass limits of $M_{*}\sim 10^{8.5} M_{\odot}$ at $z=0$ to $10^{10} M_{\odot}$ at  $z=0.75$ and \citet{Dror09}, with mass limits of $M_{*}\sim 10^{8.5} M_{\odot}$ at $z=0.2$ to $10^{9} M_{\odot}$ at $z=1$.

For $z=1-1.5$ our stellar mass functions are complete down to $M_{*}\sim 10^{8.5} M_{\odot}$, and for $z=3-3.5$ we find we are complete to $M_{*}\sim 10^{9.5} M_{\odot}$. Other work, which probes a similar redshift range, does not probe the faint end of the mass function as low as we do here. For example \citet{Pere08} has a mass limit of $M_{*}\sim 10^{9.5} M_{\odot}$ at $z=1-1.3$ and the mass limits of \citet{Kaji09} range from $M_{*}\sim 10^{9} M_{\odot}$ to $10^{10} M_{\odot}$ over $z=1-3.5$. We are thus able to examine stellar mass evolution at $z>1$ as a function of mass for the first time.

\subsection{The High Mass Galaxies and Downsizing}
In all of the stellar mass functions presented in this work, we find that the high mass galaxies have formed earlier than the low mass galaxies. For the total galaxy stellar mass functions, as shown in Figure \ref{totMF}, galaxies whose stellar mass is  $M_{*}>10^{11.5} M_{\odot}$ are close to the local value as early as $z=3-3.5$ (for $M_{*}\sim 10^{11.5} M_{\odot}$ the number densities are $31^{+13}_{-14}$ \% of the local value at this redshift). These galaxies have reached the local number density by $z=1-1.5$ ($94^{+6}_{-39}$\% the local value). Downsizing conventionally says that the high mass galaxies have stopped star forming before the low mass galaxies. However what we see here is galaxy ``mass downsizing", where by the mass of the high mass galaxy is formed before that in the low mass galaxies. This is occurring at $z>3$ unlike star formation downsizing which starts at $z<1.5$ (Bauer et al. 2010). This means the majority of the stellar mass of a galaxy is already in place before the star formation largely stops. This suggests that it is not the star formation making the galaxy massive, instead it is some other process such as mergers, which we do see happening often at high redshift (\citealt{Cons03}, \citealt{Cons06} and \citealt{Cons08}). This mass downsizing effect is also seen for the halos of galaxies (\citealt {Fouc10}).

Figure \ref{totresid} also shows the relative ratio between the local galaxy stellar mass function of \citet{Cole01} at $z=0$ and the number densities calculated in this work. The green horizonal line is how the ratio would appear if all the number densities were at the local value. The black line is a linear fit to the points from the lowest to highest ratio, hence its steepness correlates to how dominant downsizing is in that redshift range. We find that although at $z=2-2.5$ the gradient of the line is very steep, there is a generally decreasing trend with lower redshift.

We find similar results for the red and blue mass functions in Figure \ref{BRMF}, with the red population of $M_{*}\sim10^{11.5} M_{\odot}$ galaxies being very close to their local number density in the range $z=2.5-3$ ($15^{+18}_{-3}$\% for galaxies in this redshift range). The evolution of the high mass blue galaxies towards the local value is slower than that of the red. We find that the blue galaxies with stellar mass $M_{*}>10^{11.5} M_{\odot}$ are mostly present by $z=1-1.5$ ($56^{+27}_{-26}$\% for $M_{*}>10^{11.5} M_{\odot}$ for galaxies with in this redshift range). This is because these blue galaxies are likely to evolve into red galaxies at some stage. We also find downsizing in the star forming and non-star forming mass function, where the star forming/ non-star forming populations behave very similarly to the blue/red populations.

\subsection{The Intermediate Mass Dip}
Figure \ref{totresid} shows various lines indicating the ratio between the local stellar mass function and the local stellar mass function with slightly varying parameters (the coloured and dashed/dotted lines). For each line only one parameter is changed at one time. We see that between $z= 1.0$ to $1.5$ galaxies in the mass range $M_{*}\sim10^{10.5} M_{\odot} - 10^{9.5} M_{\odot}$, have generally higher ratios, hence they have generally lower number densities relative to galaxies with $M_{*}>10^{11} M_{\odot}$ and $M_{*}<10^{9.5} M_{\odot}$ . This dip feature also seems to be present for $z>2$, but is much weaker and shifted to the stellar mass range $M_{*}\sim10^{11.5} M_{\odot} - 10^{10.5} M_{\odot}$. The various coloured lines show the features of the mass function is not just an effect of the changing of the shape of the Schechter function.

The dip in the total stellar mass function has been seen before and can be explained by the differential evolution of the blue and red galaxy populations. This is consistent with studies such as \citet{Pozz09} and \citet{Ilbe09} who find a dip in the total stellar mass function that can be fit by the combination of these two population. We also find this to be the case here, as shown in Figure \ref{totcomp}. We find that a double Schechter function (the black dash dotted line), computed using the parameters of the blue and red fits, matches the form of the total Schechter function for all galaxies. The only large discrepancy occurs at the high mass end in the range $2<z<2.5$, and this corresponds with a poor fit to the red mass function.

\citet{Dror09} find that out to a redshift of $z=1$, a dip in the stellar mass function is observed not only in the total but also in the red and the blue stellar mass functions for $M_{*}\sim10^{10} M_{\odot}$ galaxies. They find that the blue stellar mass functions show a dip feature that appears to become more pronounced between $z=0.2$ and $z=1$. The also observe a dip in the red mass function that shows the opposite trend, becoming more pronounced at lower redshift. The fact that we do not see this feature in either the blue or the red mass functions tentatively suggests that any causes of the dip are perhaps not present beyond $z=1$ for these colour selected populations.

\subsection{The Low Mass Slope and $M^{*}$}
Figure \ref{totparams} shows a general steepening of $\alpha$ from low to high redshift. It is one possibility that the low stellar mass galaxies are formed early on, then undergo mergers where we know that these merger events happen at an early epoch (\citealt{Cons03}, \citealt{Cons06} and \citealt{Cons08}). These mergers form higher stellar mass galaxies and hence the the number densities of lower mass galaxies would decrease over time, causing $\alpha$ to become less steep, as is observed. They could also fade, creating a less steep mass function.

In this work we generally find a steeper $\alpha$ than previous investigations of the stellar mass function at high redshift. For example Figure \ref{totparams} shows we are steeper than both \citet{Pere08} and \citet{Elsn08}. We also find a steeper slope than \citet{Font04} ($\alpha=-1.36$ over $z=1-2$) and \citet{Font06} ($\alpha\sim -1.3$ at $z\sim 1.15$ then steepens to $\alpha\sim -1.5$ at $z\sim 3.5$). In this case, the higher stellar mass densities discussed in Section \ref{sec:massden} are most likely the result of the steeper slope. The depth of the GNS allows us to probe deeper into the low stellar mass region, allowing us to uncover the steeper slope and higher stellar mass density. We test this by recalculating stellar mass densities with less steep values for the parameter $\alpha$, but keeping the remaining parameters the same as found in this work. We find that by increasing $\alpha$ by 0.1 in the redshift range $z=2.0-2.5$ there is a $\sim 16\%$ decrease in the integrated stellar mass density. This decrease becomes  $\sim 27\%$ when alpha is increased by 0.2 in the same redshift range. \citet{Pere08} find $\alpha=-1.26$ in the redshift range $z=2.0-2.5$ and hence compute a stellar mass density of $7.87\pm 0.09$. Using this value of $\alpha$ we compute a stellar mass density of $8.06\pm 0.19$, which is within error of the value found by \citet{Pere08}. Since we can reproduce the lower stellar mass densities with a shallower value of $\alpha$, our higher stellar mass densities are likely a result of our steeper $\alpha$, and additional mass we see at the low mass end of the stellar mass function. We find that the steepness, and hence the additional stellar mass we see, is dominated by the low stellar mass blue galaxies we find. In Section \ref{sec:err} we test the effect of the errors on our number densities and find we can still reconstruct the Schechter parameters well, thus errors calculated in this work are ruled out as being the cause of our steeper $\alpha$.

For the blue mass functions in Figure \ref{BRMF}, $\alpha$ is consistent with being constant across the whole redshift range. There is also very little evolution of the parameter $\phi^{*}$, and this leads to the relatively unchanging blue stellar mass density seen in Figure \ref{SMDbr}. This is consistent with \citet{Ilbe09} who show that between $z=0.2$ and $2$ both the stellar mass function and the stellar mass density show very little evolution. This is also in agreement with \citet{Verg08} who shows the parameters for the blue galaxies are virtually unchanged between $z=0.5$ and $1.3$. This suggests there is some process by which this stellar mass function is being replenished.

We find that for the total mass function, the parameter $M^{*}$ stays roughly constant over time. This implies that the process which is increasing the numbers of galaxies at the high mass end, is doing so over the whole range of high masses above and around $M^{*}$. However this is not seen with the blue galaxies, thus this effect is not part of the evolution of this population. 

\begin{figure*}
\centering
\includegraphics[trim = 15mm 55mm 0mm 3mm, clip,scale=1.1]{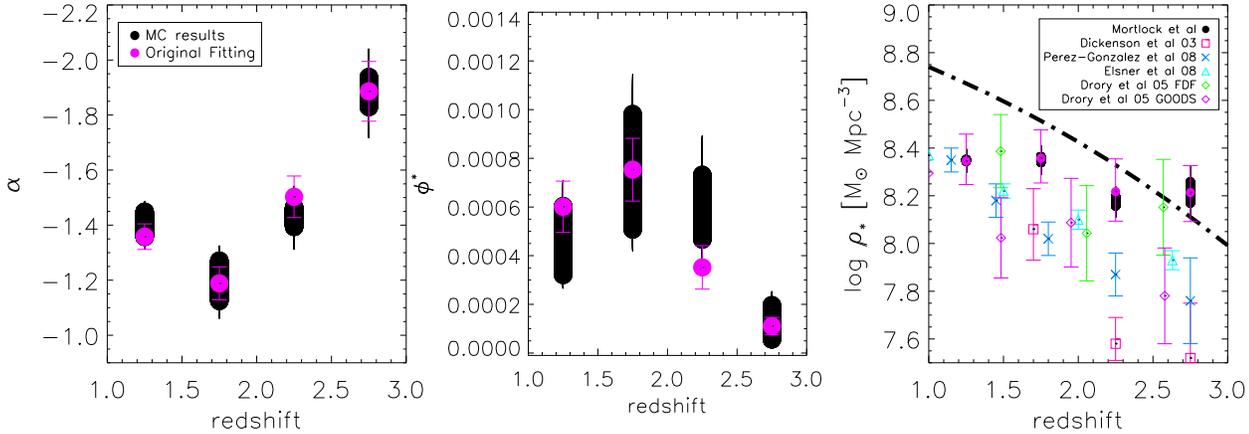}
\caption{The results of the Monte Carlo simulation to test the validity of the assumption of a constant $M^{*}$. The black circles are the parameters $\alpha$ and $\phi$, and the stellar mass densities found from the range of $M^{*}$. The pink circles show the original $\alpha$ and $\phi$ and stellar mass density from the fitting described in Section \ref{sec:fitting}. The right hand panel also includes results from literature.}
\label{mcparam}
\end{figure*}

There will be some degeneracy between the Schechter function parameters, and hence by holding $M^{*}$ constant we may be affecting the results of $\alpha$ and $\phi$. To test the validity of the assumption of a constant $M^{*}$ we use a Monte Carlo simulation to see how the parameters  $\alpha$ and $\phi$ vary with $M^{*}$. To do this we took the values of $M^{*}$ we obtained from the first fitting described in Section \ref{sec:fitting} and then varied these values between the extremes of the error on $M^{*}$. This gave us a range of $M^{*}$ and we then repeated the fitting for each value in the range keeping $\alpha$ and $\phi$ free, thus showing us how the other parameters vary. We also recalculated the stellar mass densities using Monte Carlo Schechter parameters, the parameters and stellar mass densities are also plotted in Figure \ref{mcparam}. We only apply this analysis to the range $z=1-3$ as the final redshift bin is mostly excluded from the analysis as described in Section \ref{sec:totalMF}.

We find that although variations in $M^{*}$ does produce a spread in the parameter results we still see the same general trends. We still produce steep values for $\alpha$, which in turn produce stellar mass densities that are higher than found in previous studies. We plotted histograms of the parameters and stellar mass densities and found that they do peak at the values found from fitting with constant $M^{*}$.

\section{Summary}
\label{sec:summ}
 We construct galaxy stellar mass functions using 8298 galaxies detected in the GOODS NICMOS Survey, a HST $H_{160}$-band imaging survey centred around the most massive galaxies in the redshift range $z=1.7$ to $2.9$. We use the stellar masses, redshifts, colours and star formation rates calculated from the H-band imaging combined with the ACS $BViz$ data to calculate stellar mass functions over this redshift range. We calculate the mass limit of the survey and find that we are probing galaxy stellar masses down to $M_{*}\sim10^{8.5} M_{\odot}$ in the lowest redshift bin, hence we are probing an unexplored region of the stellar mass function whilst maintaining statistically meaningful results ($\sim$300 galaxies between $z=3-3.5$). We correct for high mass bias, consider both redshift and stellar mass errors, and fit Schechter functions to examine the evolution of the parameters. We also measure the galaxy stellar mass functions for a blue/red selection  and one based on star forming/non-star forming galaxy populations. We also calculate the corresponding total stellar mass densities by integrating over the stellar mass functions at each redshift range.

The major results of this paper are as follows:
\begin{itemize}
\item We observe stellar mass downsizing in all of our stellar mass functions. For the total galaxy stellar mass functions we see that galaxies at $M_{*}>10^{11} M_{\odot}$ have reached nearly the local value by $z=3$. Even in the highest redshift bin it is clear that the most massive galaxies are very close to being in place. The same is true for the red and blue galaxies.
\item At all redshifts the blue galaxies dominate the low mass end of the stellar mass functions. Between $z=1-3$ the lowest stellar mass blue galaxies are close to the local value.
\item The total stellar mass function shows a dip feature in the intermediate mass range, which can be explained by the differential red and blue populations evolution. We see no such dip in the blue and the red stellar mass functions unlike some previous studies who have found this below $z\sim 1$. We suggest these features are not present at high redshift.

\item We find a generally steeper low mass slope for the total stellar mass functions than previous work due to the low stellar mass blue population that is probed by our deep data set. This results in a generally higher stellar mass density.
\item We find a generally higher stellar mass density, due to probing deeper into the lower stellar masses.
\end{itemize}

By constructing the galaxy stellar mass function we are investigating the mass differentiation in galaxy formation, and the way in which mass drives evolution. More work must be done to fully understand the similarities and differences between the formation of the high mass galaxies that form first, the intermediate mass galaxies who seem to show a decreased rate of formation, and low mass galaxies who are dominated by the blue population. Large surveys such as CANDELS and future spectroscopic surveys that could be performed with telescopes such as JWST and E-ELT, will provide the quality of data required to continue advancing in this field.

\section*{Acknowledgments}
We would like to thank the GNS team for their support and work on the survey and this paper as well as the referee for their helpful comments. We would also like to acknowledge funding from the STFC.

\label{lastpage}
\end{document}